\documentclass[aps,prd,twocolumn,showpacs,superscriptaddress,groupedaddress]{revtex4-1}  % for review and submission
\usepackage{graphicx}  % needed for figures
\usepackage{dcolumn}   % needed for some tables
\usepackage{bm}        % for math
\usepackage{amssymb}   % for math
\usepackage{amsmath}   % for math

\usepackage{multirow}
\usepackage{enumerate}
\usepackage{comment}
\usepackage{tabularx}
\usepackage{booktabs}

\newcommand{\no}{\mathrm}

\begin{document}

% The following information is for internal review, please remove them for submission
%\widetext
%\leftline{Version 01 as of \today}
%\centerline{\em CONFIDENTIAL}

% the following line is for submission, including submission to the arXiv!!
%\hspace{5.2in} \mbox{Fermilab-Pub-04/xxx-E}

\title{ Estimation of losses in a 300 m filter cavity and quantum noise reduction in the KAGRA gravitational-wave detector}

\author{Eleonora Capocasa$^{1,2}$} 
\email[Corresponding author:]{capocasa@apc.univ-paris7.fr}
\author{Matteo Barsuglia$^{1}$}
\author{J\'er\^{o}me Degallaix$^{3}$} 
\author{Laurent Pinard$^{3}$} 
\author{Nicolas Straniero$^{3}$} 
\author{Roman Schnabel $^{4}$} 
\author{Kentaro Somiya$^{5}$} 
\author{Yoichi Aso$^{2}$} 
\author{Daisuke Tatsumi$^{2}$} 
\author{Raffaele Flaminio$^{2}$} 

\address{$^{1}$APC, AstroParticule et Cosmologie, Universit\'e Paris Diderot, CNRS/IN2P3, CEA/Irfu, Observatoire de Paris, Sorbonne Paris Cit\'e, 10, rue Alice Domon et L\'eonie Duquet, 75205 Paris Cedex 13, France}
\address{$^{2}$National Astronomical Observatory of Japan, 2-21-1 Osawa, Mitaka, Tokyo, 181-8588, Japan}
\address{$^{3}$Laboratoire des Mat\'eriaux Avanc\'es, CNRS-IN2P3, Universit\'e de Lyon, Villeurbanne, France}
\address{$^{4}$Institut f\"{u}r Laserphysik und Zentrum fur Optische Quantentechnologien der Universitat Hamburg, Luruper Chaussee 149, 22761 Hamburg, Germany}
\address{$^{5}$Graduate School of Science and Technology, Tokyo Institute of Technology, 2-12-1 Oh-okayama, Meguro, Tokyo, 152-8551, Japan}
\date{\today}

% ---------------------------------------------------------

\begin{abstract} 

The sensitivity of the gravitational-wave detector KAGRA, presently under construction, will be limited by quantum noise in a large fraction of its spectrum. The most promising technique to increase the detector sensitivity is the injection of squeezed states of light, where the squeezing angle is dynamically rotated by a Fabry-P\'erot filter cavity. One of the main issues in the filter cavity design and realization are the optical losses due to the mirror surface imperfections. In this work we present a study of the specifications for the mirrors to be used in a 300 m filter cavity for the KAGRA detector. A prototype of the cavity will be constructed at the National Astronomical Observatory of Japan (NAOJ), inside the infrastructure of the former TAMA interferometer. We also discuss the potential improvement of the KAGRA sensitivity, based on a model of various realistic sources of losses and their influence on the squeezing amplitude. 

\end{abstract}

%\ocis{000.0000, 999.9999.}
%\pacs{04.80.Nn, 95.55.Ym, 95.75.Kk}
\maketitle

\section{\label{sec:introduction}Introduction}

Quantum noise is one of the main limiting factors for second generation gravitational-wave interferometric detectors, such as KAGRA \cite{KAGRA}, Advanced LIGO \cite {aLIGO_CQG} and Advanced Virgo \cite {AdV_CQG}. As pointed out by Caves in 1981 \cite{Caves}, the two manifestations of quantum noise, shot noise and radiation pressure noise, are created by vacuum fluctuations entering the interferometer by the antisymmetric (or \textit{dark}) port. Caves also proposed the injection of a \textit{squeezed} vacuum from the dark port as a strategy to decrease quantum noise without modifying the interferometer configuration. 

In a squeezed vacuum state, the amplitude and phase uncertainty, equally distributed in an ordinary vacuum, are modified in order to reduce one at the expense of the other.  If this is a pure state, it can be represented as an ellipse in the quadrature plane and it is characterized by the ratio of its axes (squeezing magnitude) and by its orientation (squeezing angle). Both of these parameters are functions of the Fourier frequency. If the quadrature with the reduced uncertainty is aligned in the gravitational wave signal, the signal to noise ratio is improved with respect to that achievable with an ordinary vacuum.
Since the optomechanical coupling of the laser light with the interferometer test masses induces a rotation of the squeezing ellipse, the injection of a frequency-independent squeezed vacuum (i.e with constant squeezing angle) will mitigate quantum noise only in the part of the spectrum where the gravitational-wave signal is aligned with quadrature with reduced uncertainty. In the case of a phase quadrature squeezing, the improvement is the same as that achievable by increasing the laser power. 

A broadband quantum noise reduction can be obtained by injecting a squeezed vacuum with an angle that varies with frequency, in such a way that the signal and reduced noise quadrature are always aligned \cite{kimble}. This \textit{frequency dependent squeezing} can be obtained by reflecting off a frequency-independent squeezed state by a detuned Fabry-P\'erot \textit{filter cavity}. The rotation of the squeezing angle has been experimentally demonstrated in the MHz region\cite{Che} and, more recently, in the kHz region\cite{Oel}. 

The injection of frequency dependent squeezing is particularly suitable for a future upgrade of KAGRA, since increasing  laser power to reduce shot noise can be difficult due to the cryogenic temperature of the detector. Moreover, since thermal noise is low enough to make quantum noise the dominant contribution in a large part of the spectrum, the reduction of quantum noise would have a direct and significant effect on the detector sensitivity.

The achievement of a high level of frequency dependent squeezing faces two kinds of difficulties. First, the squeezing angle should undergo a rotation in the frequency region where quantum noise switches from radiation pressure noise to shot noise ($\sim 70$ Hz for KAGRA). This corresponds to a filter cavity with a very long storage time of $\sim 2$ ms. Moreover, the presence of optical losses in the filter cavity system will decrease the squeezing factor, since losses are associated with ordinary vacuum fluctuations which couple with the squeezed states reducing their squeezing level. For both these reasons, optical losses in the filter cavity system should be estimated and reduced as much as possible. 

A 300 m prototype of the cavity will be developed at NAOJ, using the former TAMA infrastructure \cite {TAMA}. This is, to our knowledge, the longest filter cavity prototype under construction. Previous experiments were made using a 2 m long cavity \cite {Oel}. The first goal of this paper is to present the filter cavity optical design. One of the main issues in the filter cavity realization are the losses due to the mirror surface imperfections. In this work we present a study of the specifications for the mirrors, based on real mirror maps, used in the Virgo experiment. Moreover, a crucial point is to compare the squeezing degradation from the filter cavity optical losses with those originated by other mechanisms. Thus, we present a complete \textit {squeezing degradation budget} for the 300 m cavity, based on the work by Kwee et al.\cite {sq2}. We finally discuss the potential improvement of  KAGRA sensitivity using the 300 m filter cavity, compatible with the KAGRA infrastructure.

The interest of this work goes beyond KAGRA, since a 100-m class filter cavity is also a possible solution for a medium-term upgrade of Advanced Virgo\cite {virgo_sqz}, and 300 m filter cavities are planned for the 3rd generation detector Einstein Telescope\cite {ET}. The structure of the article is as follows: in section \ref{sec:fc design} we demonstrate the computation of the filter cavity parameters needed to obtain an optimal rotation of the squeezing angle for KAGRA. 
In section \ref{sec:simulation} the filter cavity numerical simulation, with realistic mirror maps, is presented.  
 %A similar study has be done in [5] for a 10 m cavity and in our analysis we used the procedure and the numerical tool developed in this contest, adapting them to a longer cavity.
In section \ref{sec:degradation} an estimation of the squeezing degradation, due to cavity losses and other degradation mechanisms, is obtained. Finally, the expected improvement in KAGRA sensitivity, achievable using frequency-dependent squeezing, is presented.

%The sensitive parameter is the round trip losses per cavity unit length \cite{Khalili1}. A preliminary study based on LIGO performance  has set the target for this value to 1ppm/m \cite{sq1}.

%A particularly suitable form to describe this kind of manipulation has been developed (caves). In this formalism, referred to as two-photon, fluctuations are described by a quadrature vector, whose elements account for amplitude and phase fluctuations as a function of side-band frequency. Rotations, squeezing and propagation are described as matrices to be applied to the quadrature vector.
%A single filter cavity is enough to provide the desired squeezing rotation only in interferometers in broadband configuration \cite {kimble}.

\section{\label{sec:fc design}Filter cavity optical design}     
%The filter cavity simulation which we performed aims to set a requirement on the mirror flatness, needed to keep  round trip losses  below a certain amount. The sensitive parameter for the filter cavity performance is the round trip losses per cavity unit length \cite{Khalili1}. A preliminary study based on LIGO performance has set the target for this value to 1ppm/m \cite{sq1}. 

In the first part of this section we discuss the choice of finesse and detuning of the filter cavity. In the second part, we address the problem of the choice of cavity length, mirror dimension and radius of curvature. 

\subsubsection{Length, finesse and detuning}

Filter cavity bandwidth and detuning should be chosen in order to allow an optimal rotation angle of the squeezed state to be injected in the interferometer dark port. This rotation should counteract that which is produced by the interferometer, keeping the squeezing state aligned with the signal quadrature at all frequencies. The optimal rotation angle of the cavity is therefore a function of the interferometer parameters \cite{Harms}:

\begin{equation}\label{tfc}                       
\theta_{\no{fc} }(\Omega)= \arctan(K(\Omega))
 \end{equation}
with $K$ being the frequency dependent optomechanical coupling, whose dependance from the signal sideband frequency $\Omega$ given by:

\begin{equation}\label{k}                       
K(\Omega) = \left(\frac{\Omega_\no{{SQL}}}{\Omega}\right)^{2} \frac{\gamma^{2}_\no{{ifo}}}{\Omega^{2}+\gamma^{2}_\no{{ifo}}}
\end{equation}

where $\gamma^{2}_{\no{ifo}}$ is the interferometer bandwidth and $\Omega_{\no{SQL}} $ is the frequency at which quantum noise reaches the \textit{standard quantum limit}\cite{Brag}, marking the transition between radiation pressure noise and shot noise. In a dual recycled interferometer, with a tuned signal-recycling cavity, we have: 

\begin {equation}
\Omega_{\no{SQL}} =\left[\frac{t_{\no{sr}}}{1+r_{\no{sr}}} \right]\frac{8}{c} \sqrt{\frac{P_{\no{arm}}\omega_{0}} {mT_{\no{arm}}} }                              
\end{equation}
and
\begin {equation}
\gamma_{\no{ifo}} =\left[\frac{1+r_{\no{sr}}}{1-r_{\no{sr}}} \right]\frac{T_{\no{arm}}c} {4L_{\no{arm}}}                               
\end{equation}

where $t_{\no{sr}}$ and $r_{\no{sr}}$ are the signal recycling mirror amplitude transmissivity and reflectivity, $P_{\no{arm}} $ is the intracavity power, $T_{\no{arm}}$ is the arm cavity input mirror power transmissivity, $\omega_{0}$ is the angular frequency of the carrier field, $m$ is the mirror mass. 

For KAGRA, using the parameters shown in Tab.  \ref{kagrap}, we have:
 
\begin{equation}                                                                                                            
\Omega_{\no{SQL}} = 2\pi \times 76.4 \, \no{Hz}
\end{equation}  

and

\begin{equation}                                                                                                            
\gamma_{\no{ifo}} = 2\pi \times 382\, \no{Hz}
\end{equation}

For $\Omega <<\gamma_{\no{ifo}}$ Eq.\ref{k} can be simplified and we are in the region of interest. The required rotation angle then becomes:

 \begin{equation}\label{tfc}                       
\theta_{\no{fc}}(\Omega)= \arctan\left(\frac{\Omega_\no{{SQL}}}{\Omega}\right)^{2}
 \end{equation}

 \begin{table}[!ht]
 \begin{center}    
 \begin{ruledtabular}
\begin{tabular}{llr}

Parameter                                                                                 & Symbol                                          & Value \\
\hline

Carrier field frequency                                                                &$ \omega_{0}$                              & $2\pi \times 282 \, \no{THz} $   \\
Standard quantum limit frequency                                              & $\Omega_{\no{SQL}}$                & $2\pi \times 76.4 \, \no{Hz} $   \\
Arm input mirror transmissivity                                                   & $T_{\no{arm}} $                           & 0.004       \\
Signal recycling input  transmissivity                                          &  $t^{2}_{\no{sr}}   $                      & 0.1536      \\
Intracavity power                                                                        &   $P_{\no{arm}} $                         & $  400  \,\no{kW} $   \\
Mirror mass                                                                                & $m$                                              & $22.8   \,  \no{kg} $  \\

 \end{tabular}
\end{ruledtabular}
\caption{Values and symbols for KAGRA interferometer parameters.}
\label{kagrap}
\end{center}
\end{table}  

In order to obtain the above frequency dependent rotation, the bandwidth $\gamma_{\no{fc}}$ and the detuning $\Delta \omega_{\no{fc} }$ of a lossless filter cavity should be:
 
 \begin{equation}\label{tfc}                       
\gamma_{\no{fc} }= \frac{\Omega_\no{{SQL}}}{\sqrt{2}}
\end{equation} 
 
\begin{equation}\label{det}                 
\Delta \omega_{\no{fc} }= \gamma_{\no{fc}}
\end{equation}

As discussed in \cite{sq2}, in presence of losses, the bandwidth and the detuning of the filter cavity become:

\begin{equation}\label{gamma2}
\gamma_{\no{fc}} = \sqrt{\frac{2}{(2-\epsilon)\sqrt{1-\epsilon}}}\frac{\Omega_{\no{SQL}}}{\sqrt{2}}
\end{equation} 

\begin{equation}\label{det}                 
\Delta \omega_{\no{fc} }= \gamma_{\no{fc}} \sqrt{1- \epsilon}
\end{equation}

where $\epsilon$ is a function of the filter cavity round trip losses $\Lambda^{2}_{rt}$, the free spectral range $f_\no{FSR} = c/2L_{\no{fc}}$, and $\Omega_{\no{SQL}}$:  

\begin{equation}\label{eps} 
\epsilon = \frac{4}{2+\sqrt{2+2\sqrt{1+\left(\frac{2\Omega_{\no{SQL}}}{f_\no{FSR} \Lambda^{2}_{rt}}\right)^{4} }}}
\end{equation}
 
Given a filter cavity with fixed length and losses, we can compute the parameter $\epsilon$ using Eq. \ref{eps}. Then, using Eq. \ref{gamma2} and \ref{det}, we obtain  the optimal bandwidth and detuning. Finally, in order to compute the cavity finesse, we can write the bandwidth $\gamma_{\no{fc}}$ of a Fabry-P\'erot  cavity in terms of the losses, length and input mirror transmissivity $t^{2}_{\no{in}}$: 

\begin{equation}\label{gamma1} 
\gamma_{\no{fc}} = \frac{t^{2}_{\no{in}}+\Lambda^{2}_{rt}}{2}f_\no{FSR}
\end{equation} 

Inverting the previous equation, we can finally compute $t^{2}_{\no{in}}$ which determines the cavity finesse.

The length of the filter cavity is set to 300 m, corresponding to the length of the arms of the former TAMA interferometer. The losses $\Lambda^{2}_{rt}$ are set to 80 ppm, a value which will be justified in the following section. For these values, we found $\epsilon =  0.111 $  and  $\gamma_{\no{fc} }= 2\pi \times 57.3 $. The optimal detuning  will be $ \Delta \omega_{\no{fc} } = 2\pi \times 54 \, \no{Hz}  $ and  $t^{2}_{\no{in}}  = 0.0014$, corresponding to a finesse of $4480$. We highlight the fact that for  $\Lambda^{2}_{rt}$ up to $ \sim 700 $ ppm, the value of the finesse is almost independent of the cavity losses. 

\subsubsection{Mirrors dimension and radius of curvature (RoC)}

In order to reduce the losses due to the finite size of the mirrors (clipping losses), we require the size of the beam to be as small as possible. As shown in \cite{fab}, this is also the best way to reduce the effects of large-scale mirror defects. The confocal configuration (radii of curvature equal to the cavity length) gives the smallest beam radius on the mirrors, but this configuration is marginally stable. We choose radii  $\sim 400$ m, reasonably larger than  300 m, to avoid cavity instability. The exact value of the RoC will be determined using numerical simulations, described in the next paragraph, in order to minimize the losses. 

%The simulation to measure losses level has been performed over a range of RoC between 360 and 420 m\
For a 300 m cavity with two mirrors with a RoC of 400 m, the beam diameter at the waist and on the mirrors is respectively 0.0162 m and 0.0205 m. Choosing, as it was done for TAMA, mirrors with a diameter of 0.1 m, (roughly five time bigger that the beam radius) results in completely negligible clipping losses.\  

%In fact, the cavity results degenerated for some values of the curvature radius. As we will see, simulations will point out such values and the radius of curvature will be picked in order to be as far as possible from them.

The filter cavity parameters are reported in Tab. \ref{recparfc}

\begin{table}[!ht]
 \begin{center}    
 \begin{ruledtabular}
\begin{tabular}{llr}

Parameter                                                                                 & Symbol                                             & Value \\
\hline
Length                                                                                        & $L$                                                & $300$ m \\
Radius of curvature                                                                      & RoC                                               & variable \\
Mirror diameter                                                                             & $d$                                                 & $0.1 $ m  \\

Input mirror transmissivity                                                            &  $t^{2}_{\no{in}}   $                       & 0.0014     \\
Finesse                                                                                         &  $F$                                               &   $4480$  \\

Beam diameter at waist(RoC 400 m)                                          &                                                      & $0.0162$ m  \\
Beam diameter at the mirror(RoC 400 m)                                   &                                                      & $0.0205$ m  \\

\end{tabular}
 \end{ruledtabular}
\caption{Filter cavity parameters. }
 \label{recparfc}
 \end{center}
\end{table}

\section{\label{sec:simulation}Mirror surface quality specifications}
The presence of losses in the filter cavity, caused by mirror defect and mirror finite size, induces squeezing degradation. In this section we will detail the procedure needed to find the specifications on mirror dimensions and surface flatness.

Mirror defects, namely deviations of the mirror surface from a perfect spherical one, are described by the \textit{mirror map}, which is a square matrix of $n$ elements with $n = l_{mir}/res$, where $l_{mir}$ is the length of the surface described, and $res$ is the resolution of the map.  Each matrix element, corresponding to a pixel with area $res^{2}$, contains a measure of the mirror surface height $h$ (with respect to a perfect spherical one).

We can associate to each mirror map (or to a part of it) the Root Mean Square (RMS) of the height, defined as:

	 \begin {equation} \label{}
	\sigma_{\no{RMS}} = \sqrt{\frac{1}{n}\sum_{i=1}^{n} (h_{i}-\bar{h})^{2}}  \qquad   \no{with} \quad\bar{h} = \sum_{i=1}^{n} h_{i} 
	\end{equation} 

Another useful number to quantify surface flatness is the peak-to-valley value (PV): a measure of the difference between the highest and the lowest point.

Mirror defects can be studied in the spatial frequency domain by applying a 2D Fourier transform to the mirror map. The lowest spatial frequency $f_{min}$ coincides with the inverse of $l_{mir}$, while the maximum spatial frequency $f_{max} $ is given by$1/(2\cdot res)$.  A 1D Power spectral density (PSD) can be associated with the 2D Fourier transform map \cite{hiro}. For such a 1D PSD we have the relation:  

	\begin {equation} \label{psd}
	\sigma^{2}_{\no{RMS}} = \int_{f_{min}}^{f_{max}} \no{PSD}(f) \,df
	\end{equation}

The low frequency defects, those which have a spatial frequency up to $10^{3} m^{-1}$, contribute to the so-called \textit{mirror flatness} while higher frequency defects are associated with the \textit{mirror roughness}.\ This distinction has no physical motivation, but it is simply due to different techniques used to measure spatial defects in the two cases \cite{nicolas}.

The scattering angle for light at normal incidence of wavelength $\lambda$ can be written as a function of the frequency of spatial defect as \cite{sto}:
 
\begin {equation} \label{theta}
\theta= \lambda \times f
\end{equation}

This equation means that a defect at spatial frequency $f$ will scatter a fraction of the light at angles $\theta$ or larger. As a consequence, the amount of light reflected back at normal incidence will be reduced by the same amount.  This fraction is given by $ (4\pi \times \sigma(f) / \lambda)^2$ where $\sigma(f)$ is the amplitude of the defect at spatial frequency $f$.

For a given cavity length $L$ and mirror diameter $d$, there is a maximum scattering angle $\theta_{\no{limit}}$ above which light is scattered out of the cavity: 

\begin {equation} \label{tmax}
\theta_{\no{limit}} = \frac{d}{2L}
\end{equation}

Using Eq. \ref{theta} and \ref{tmax} we can then find a spatial frequency  $f_{\no{limit}}$ for the mirror defects above which the light is scattered out of the cavity:

\begin{equation} \label{f_lim}
f_{\no{limit}} = \frac{d}{2L\times \lambda}
\end{equation} 

From the equation above, the losses due to defects with spatial frequency above $f_{\no{limit}}$ can be estimated as \cite{sto}:

	\begin{equation}\label{raflos1}
	\no{losses}_{(f>f_{\no{limit}})} = \left(\frac{4\pi \times \sigma}{\lambda}\right)^{2}
	\end{equation} 

where $\sigma$ is the RMS for frequencies above $f_{\no{limit}}$. For the filter cavity we are considering $f_{\no{limit}}=157 m^{-1} $. 

It is important to note that the RMS for frequencies lower than $f_{\no{limit}}$ also contributes to losses. In fact, even if light is not immediately scattered out of the cavity, it is likely to be transferred on higher order modes and eventually exit the cavity. An accurate estimation of the round trip losses has been done using a numerical simulation of the fields in the cavity after adding realistic maps to the mirror surfaces. This allowed us to set specifications for the surface defects flatness needed to keep losses below a desired threshold. 

The round trip losses in a Fabry-P\'erot cavity are defined as \cite{rtl_ref}:

\begin {equation} \label{RTL}
\Lambda^{2}_{rt} = \frac{P_{in}-P_{r}-P_{t}}{P_{circ}}
\end{equation} 

where $P_{in}$ is the input power (which is assumed to be a fundamental mode), $P_{circ}$, $P_{t}$, $P_{r}$ are the powers circulating in the cavity, transmitted and reflected, respectively. Since we can only take advantage of the light reflected on the fundamental mode, the definition of Eq. \ref{RTL} has been modified to:

\begin {equation} \label{RTL00}
\Lambda^{2}_{rt} = \frac{P_{in}-P^{00}_{r}}{P_{circ}}
\end{equation} 
where $P^{00}_{r}$ is fraction of the reflected power which is on the fundamental mode.
%In addition to light which exit the cavity because of the finite dimension of mirror, the presence of defects on mirror surface is responsible for scattering light out of the %cavity or to transfer it from the fundamental mode to higher order mode. 

We used  the MATLAB package OSCAR \cite {oscar} to simulate the fields and compute the values of $P_{circ}$  and $P^{00}_{r}$ to be used in Eq.\ref {RTL00}.

\begin{figure*}
	\centering{\includegraphics[width=0.8\textwidth]{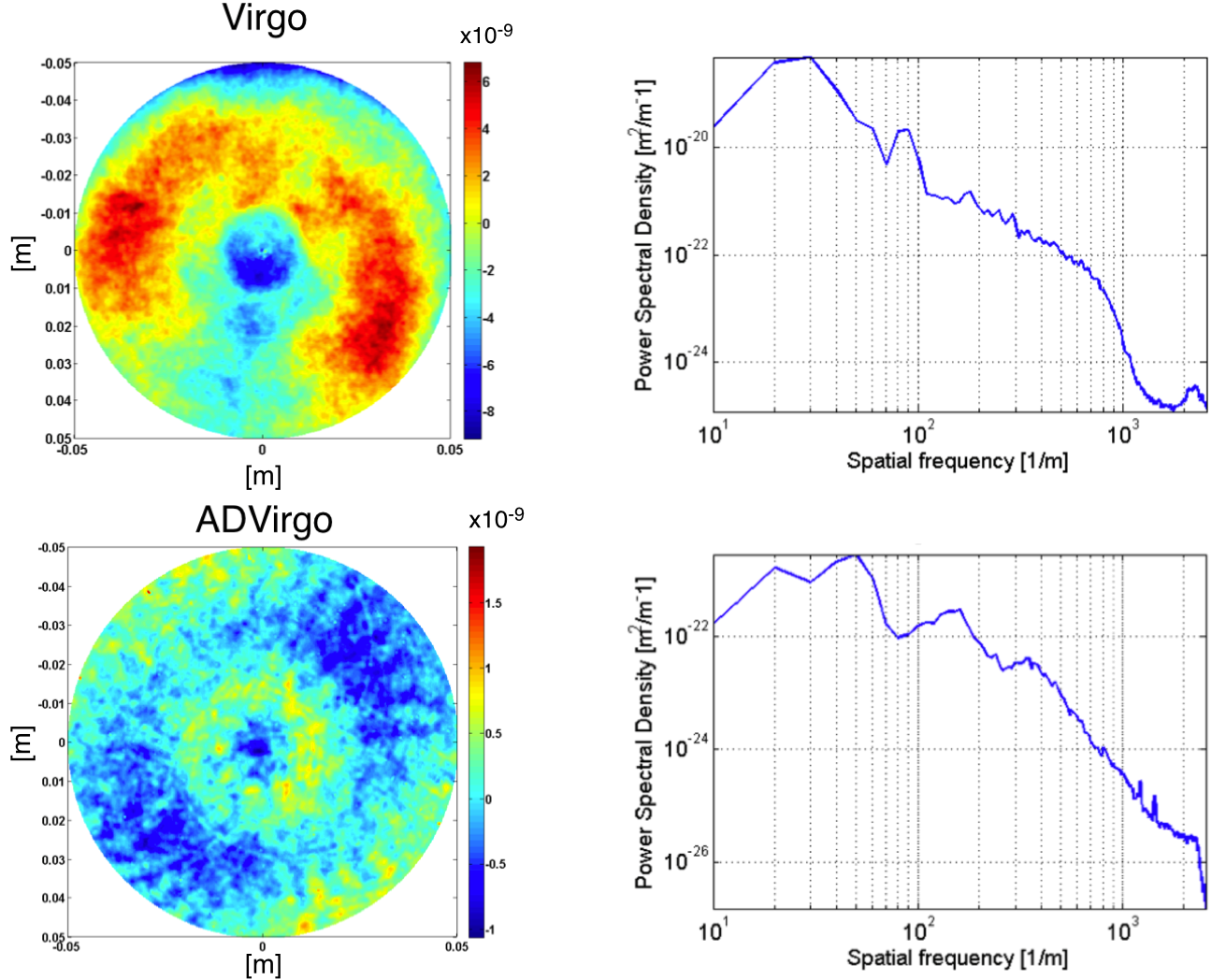}}
	\caption{Initial Virgo map (top) and Advanced Virgo map (bottom). Mirrors maps (left), PSD (right).}
	\label{maps}
\end{figure*}

We ran the simulation using five different mirror maps, for mirrors used in the Virgo experiment. Four of these mirrors were produced for the initial Virgo experiment (2007-2011) with a standard polishing technology. The fifth belongs to the Advanced Virgo experiment (which will become operational in 2016) and is obtained with an ion beam polishing technique. The Virgo maps have a resolution of about $350\, \no{\mu}$m, while the Advanced Virgo map has a resolution of $378.4\, \no{\mu}$m. One example of the Virgo maps and the Advanced Virgo map are shown in Fig. \ref{maps} along with their relative PSD.  %\footnote{http://lma.in2p3.fr/Activites/databasegb.htm}. 

The cavity parameters used in the simulation are those reported in Tab. \ref {recparfc}. The maps were only applied to the end mirror, while the input mirror has been considered perfect. We checked that the round trip losses for a cavity where both the mirrors have defects can be obtained by simply multiplying by two the previous result. Results presented here have already been multiplied by two.

\begin{figure}
	\centering{\includegraphics[width=0.5\textwidth]{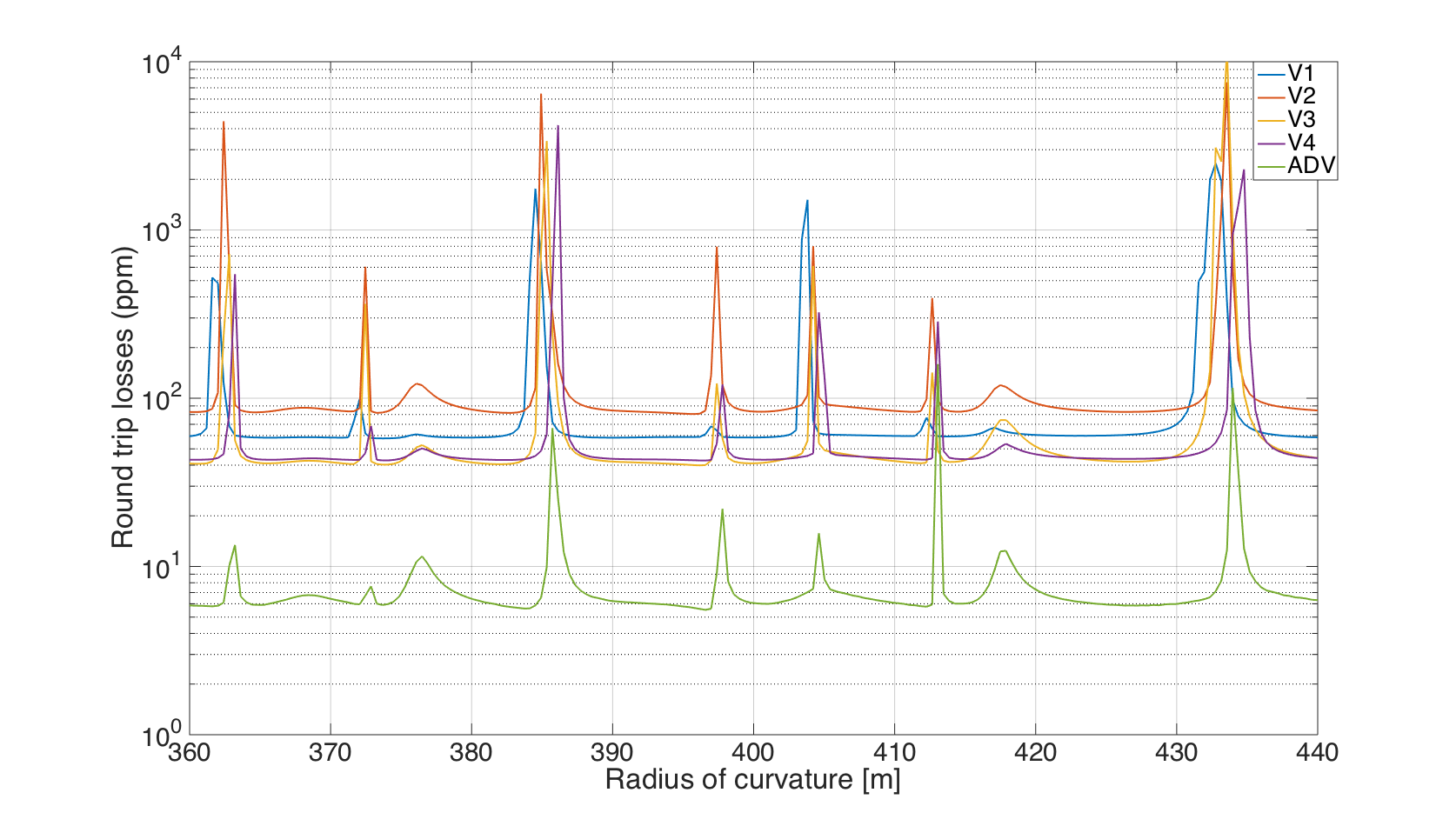}}
	\caption{Round trip losses as a function of the radius of curvature. Peaks correspond to values of RoC for which the cavity is quasi-degenerate.}
	\label{recapfig}
\end{figure}

Each surface has been characterized by its RMS and its PV over different diameters. The measured values are reported in Tab. \ref{recap}. Round trip losses for the various mirrors have been calculated in terms of the radius of curvature and are reported in Fig. \ref{recapfig}. The losses floor for each mirror has also been reported in the last column of Tab. \ref {recap} in order to be directly compared with the mirror flatness. The presence of peaks in the plots of Fig.\ref{recapfig} is due to power transferred to higher order modes which are partially resonant along with the fundamental mode for certain values of the curvature radius. 

\begin{figure*}
	\centering{\includegraphics[width=0.7\textwidth]{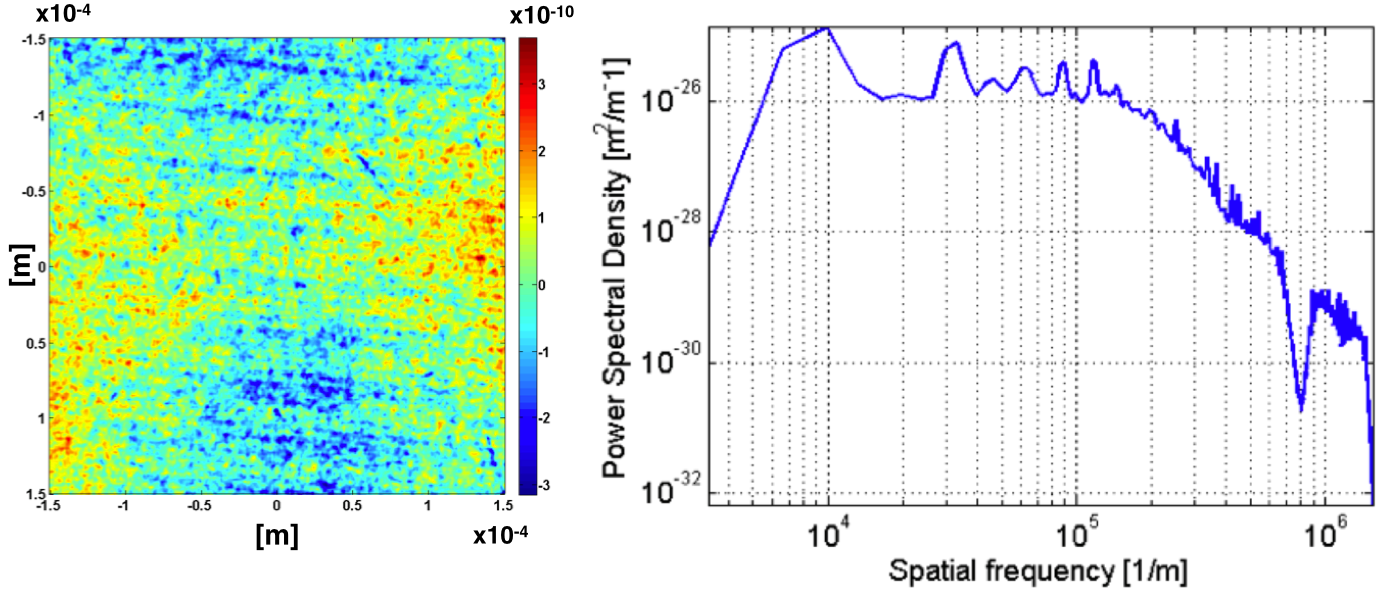}}
	\caption{Roughness maps for the Advanced Virgo configuration. Mirror map (left), PSD (right).}
	\label{micromap}
\end{figure*}

Maps used in the simulation account for mirror defects with spatial frequency going from 10 $m^{-1}$ to $2\cdot10^{3} m^{-1}$. This means that losses caused by mirror roughness are not included in this estimation. A map of the the roughness has been measured for the Advanced Virgo mirror.  this map, obtained by scanning an area of  0.3 mm x 0.3 mm with a resolution of $1.28\, \no{\mu}$m, scans frequencies from $3.3\cdot10^{3}$ to $3.9\cdot10^{5} m^{-1}$. From its RMS, under the assumption that it is uniform on the surface, we can estimate additional losses due to roughness, which should be added to those already measured. The RMS is 0.08 nm, which corresponds to 0.89 ppm of additional losses for a single reflection. For the initial Virgo mirrors,  the specification on the roughness RMS was  0.1 nm which corresponds to 1.4 ppm of additional losses for a single reflection. In both cases the losses are dominated by flatness defects.The roughness map for Advanced Virgo and the relative PSD are shown in Fig.\ref {micromap}. 
The measurement of the light scattered at angles larger than a few degrees gives losses of the order of 5 ppm. These include both the losses due to the roughness discussed above and those due to point defects. The corresponding additional round-trip losses will be about 10 ppm. We remark that, even including all these effects, still difference exists between the measured losses in Advanced LIGO and simulation results. Losses due to scattering at angles between mrad and a few degrees are being investigated as a possible cause of these differences\cite{hiroprivate}. These hypothetical losses are not included in this paper.

The conclusions of this study are that, for his filter cavity (length $300$ m, mirror diameter $10$ cm, RoC $\sim 400$ m):

\begin {itemize}

\item An Advanced Virgo-class mirror will produce floor losses $< 10$ ppm and a Virgo-class mirror will produce losses $\sim 40-80$ ppm. In order to determine the final specifications on the mirror flatness, the squeezing degradation given by the cavity losses has to be compared to the other degradation mechanisms. This analysis is performed in the next section. 

\item The accidental degeneracy can amplify the losses by more than an order of magnitude. The simulation gives the \textit{safe} regions, where the losses are at the floor level.

\item The precise RoC value should be chosen in some of the floor losses regions. A precision on the RoC of $\sim 1 \%$ is necessary to guarantee the RoC to be in these regions. 

\end{itemize}

\begin{table*}[]
\centering
\label{recap}
%begin{ruledtabular}
%\end{ruledtabular}
\scalebox{1.2} {
\begin{tabular}{|l|l|l|l|l|l|l|l|l|l|}
\hline
     & \multicolumn{2}{c|}{\begin{tabular}[c]{@{}c@{}}diameter\\  0.10 m\end{tabular}}                       & \multicolumn{2}{c|}{\begin{tabular}[c]{@{}c@{}}diameter\\ 0.05 m\end{tabular}}                                              & \multicolumn{2}{c|}{\begin{tabular}[c]{@{}c@{}}diameter\\ 0.02 m\end{tabular}}                                              & \multicolumn{2}{c|}{\begin{tabular}[c]{@{}c@{}}diameter\\  0.01 m\end{tabular}}                                             &                                                               \\ \hline
Mirror & \begin{tabular}[c]{@{}c@{}}RMS\\ (nm)\end{tabular} & \begin{tabular}[c]{@{}c@{}}PV \\ (nm)\end{tabular} & \multicolumn{1}{l|}{\begin{tabular}[c]{@{}l@{}}RMS\\ (nm)\end{tabular}} & \begin{tabular}[c]{@{}c@{}}PV\\ (nm)\end{tabular} & \multicolumn{1}{l|}{\begin{tabular}[c]{@{}l@{}}RMS\\ (nm)\end{tabular}} & \begin{tabular}[c]{@{}c@{}}PV\\ (nm)\end{tabular} & \multicolumn{1}{l|}{\begin{tabular}[c]{@{}l@{}}RMS\\ (nm)\end{tabular}} & \begin{tabular}[c]{@{}c@{}}PV\\ (nm)\end{tabular} & \textbf{\begin{tabular}[c]{@{}c@{}}Losses\\ (ppm)\end{tabular}} \\ \hline
V1     & 2.617                                            & 15.95                                              & 1.424                                                                   & 9.46                                              & 0.687                                                                   & 6.04                                              & 0.558                                                                   & 5.38                                              & 57.8                                                            \\ \hline
V2     & 1.875                                            & 15.64                                              & 1.234                                                                   & 8.56                                              & 0.682                                                                   & 6.29                                              & 0.812                                                                   & 5.92                                              & 80.6                                                            \\ \hline
V3     & 2.499                                            & 15.34                                              & 1.360                                                                   & 10.51                                             & 0.754                                                                   & 4.31                                              & 0.430                                                                   & 3.31                                              & 39.8                                                            \\ \hline
V4     & 1.752                                            & 45.61                                              & 0.984                                                                   & 12.12                                             & 0.509                                                                   & 4.46                                              & 0.531                                                                   & 4.46                                              & 42.6                                                            \\ \hline
ADV    & 0.319                                            & 2.99                                               & 0.274                                                                   & 2.09                                              & 0.192                                                                   & 1.18                                              & 0.142                                                                   & 0.97                                              & 5.6                                                             \\ \hline
\end{tabular}}
\caption{RMS and  PV (over different diameters) and the round trip losses floor for each mirror map. The values indicated for the losses correspond to the floor of Fig.\ref{var_diam}.}
\end{table*}
%Tilt and focus have been removed over the diameter where RMS and PV are calculated. 
%For this reason it is possible sometimes to find a higher value for PV and RMS when restricting the diameter (See  Virgo mirror VEM 10).
 %The round trip losses in the fundamental mode has been compared with the total losses, obtained from \ref{RTL00} where this time $P_{r}$ takes into account all the reflected power.

%\begin{figure}
%	\centering{\includegraphics[width=0.5\textwidth]{recapfig1.png}}
%	\caption{}
%	\label{recapfig1}
%\end{figure}

\subsubsection*{Mirror dimensions}

\begin{figure}[htb]
	\centering{\includegraphics[width=0.5\textwidth]{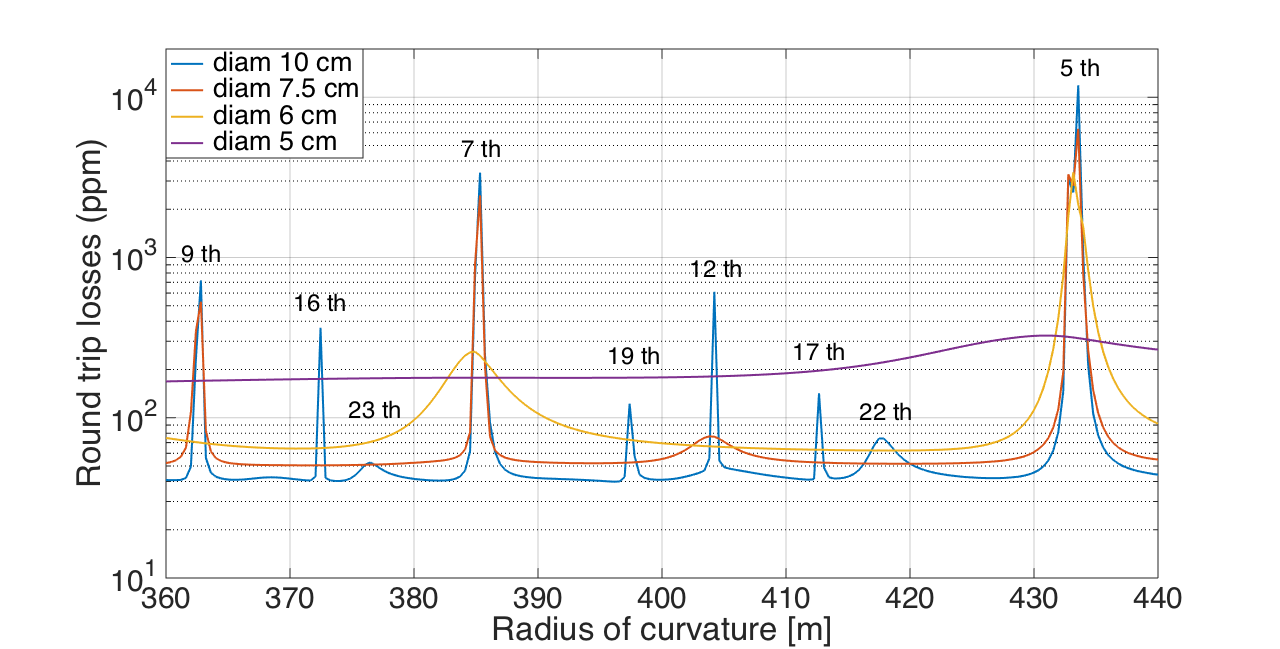}}
	\caption{Round trip losses as a function of RoC for different values of the mirror diameter. Peaks which correspond to cavity degeneracies are reduced for smaller mirrors while the  floor  of round trip losses increases.}
	\label{var_diam}
\end{figure}
Choosing the best mirror dimension is not as straightforward as it may seem to be. The scattering mechanisms described in the previous paragraph show that bigger mirrors have lower levels of losses. In fact, we see from Eq.\ref{f_lim} that by increasing the  mirror diameter $d$, the minimum spatial frequency $f_{limit}$ of defects which scatter light out of the cavity is higher. Consequently, the RMS for frequencies above $f_{limit}$ in Eq.\ref{raflos1} is reduced and so too the losses. However this is not the only effect to be taken into account. We have already observed that for certains values of the curvature radius, the cavity can be degenerate. This means that the separation between the resonance frequency of the fundamental mode and that of a higher order mode is small enough to make it partially resonant. Some power of the fundamental mode is then transferred to this higher order mode, and losses in the fundamental mode are remarkably increased. Critical RoC values are highlighted by simulation and correspond  to the peak observed in Fig.\ref{recapfig}. Mirror dimensions also have a strong impact on the appearance of such peaks. In Fig.\ref{var_diam}, round trip losses for a virgo mirror (V3) are shown as a function of the RoC for different values of the mirror diameter. We see that for bigger diameters more peaks are observed, i.e. there are more RoC values which make the cavity degenerated. This number is reduced for smaller mirrors. This effect can be possibly explained by considering the intensity profile of higher order modes. In fact, the power of higher mode is spread on a bigger surface than the fundamental mode. Generally, the \textit{width} of a mode increases with its order, and then, a reduction of the mirror dimension prevents wider modes from resonating. This  explanation is confirmed by the fact that a gradual reduction of  mirror dimension first eliminates resonances of the modes with higher $n$. 
Using smaller mirror will reduce the number of peaks, while increasing the floor losses level according to the mechanism explained at the beginning of the paragraph. Therefore mirror dimensions should be chosen in order to strike the best balance between low floor losses and the presence of a safe zone for the RoC reasonably far from degeneracies. Using smaller mirrors allowed us to relax accuracy requirements on the RoC value. We can see from Fig. \ref{var_diam} that a mirror diameter of 0.075 m allows  us to pick a RoC, for example of 420, for which losses remain basically constant in a range of $\sim 20$ m. Therefore the required precision on the RoC is  $\sim 5 \%$. For such a mirror dimension the contribution of clipping losses to cavity losses are still negligible.
An initial value for the mirror dimension has been set to 0.1 m. This was the diameter of the former TAMA mirrors and it will allow the reuse of part of the suspension system already developed for TAMA. If needed, diaphragms can be placed in front of the mirror to reduce their diameter according to what we have observed.

\subsubsection*{Simple round trip simulation}
\begin{figure}[htb]
	\centering{\includegraphics[width=0.3\textwidth]{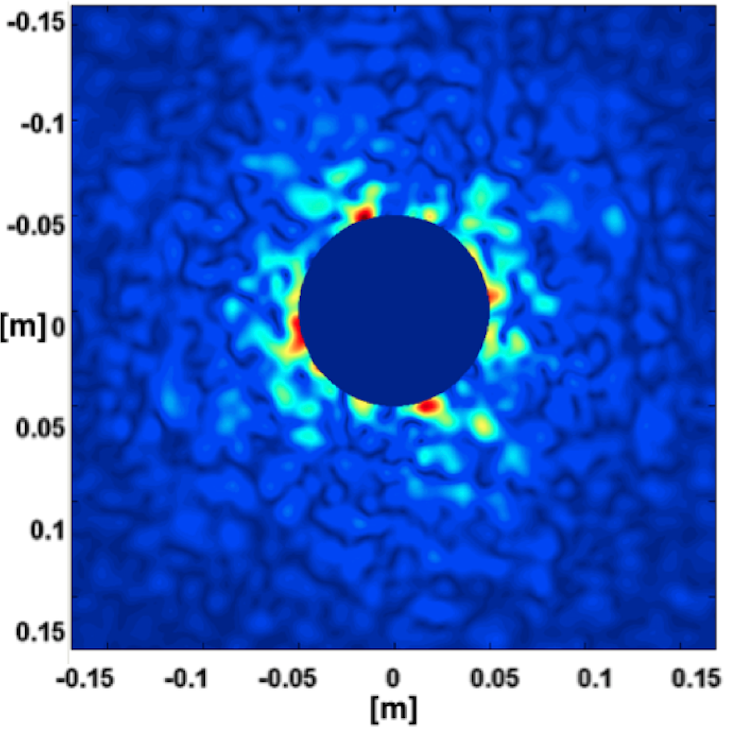}}
	\caption{Amplitude of the field scattered out of the cavity by high spatial frequency defects. The field is observed after being reflected from a mirror with a real map and then propagated for 300 m. }
	\label{eclissi}
\end{figure}

A complete simulation of the cavity has been used to compute fields in our system and then to evaluate round trip losses, according to Eq.\ref{RTL00}. This simulation allows us to take into account the effect of higher order mode resonance. In this case, light does not exit the cavity but is partially transferred to a higher order mode. This effect is observable by comparing the round trip losses on the fundamental mode with those in all modes (i.e the total amount of light exiting the cavity).%plotted in figure?
Coinciding with resonance,  the curves relative to the two cases show a discrepancy accounting for the light which is still in the cavity but not on the fundamental mode.
On the other hand, when the cavity is not degenerate, the total amount of light exiting the cavity coincides with the light lost on the fundamental mode. This quantity corresponds to the floor losses level reported in Tab. \ref {cfr_cq} and has been compared with that obtained by simply computing the power escaping the mirror aperture after a single roundtrip (reflection from a mirror with real map, propagation for 300 m, then reflection again from a perfect mirror (as in the simulated cavity) and propagating again for 300 m).
In practice, we measure the lost power after a cavity round trip of the beam. The comparison between losses obtained with the complete simulation and those found with the simple round trip are shown in Tab. \ref{cfr_cq} for different mirrors maps and for different values of the diameter and shows a good agreement. Even if the second method is less accurate and cannot be used in the case of degeneracy, it is much faster and can be employed to obtain a rough estimation of the floor level of the round trip losses  in the cavity. 
A plot of the light power exiting cavity after the first reflection is shown in Fig. \ref{eclissi}.

\begin{table}[!h]
 \begin{center}    
 \begin{ruledtabular}
 \begin{tabular}{cccc}
\multicolumn{1}{l}{Mirror} & Diameter & RTL - full sim & RTL - quick sim\\
\hline
V1                         & 10 cm    & 57.8 ppm           & 57.6 ppm            \\
V2                         & 10 cm    & 80.6  ppm         & 79.4 ppm            \\
V4                         & 10 cm    & 42.6  ppm         & 40.4 ppm            \\
ADV                      & 10 cm    & 5.6  ppm           & 5.6 ppm            \\
V3                         & 10 cm    & 39.8  ppm         & 38.8 ppm            \\
V3                         & 7.5 cm   & 50.4 ppm          & 50.0 ppm            \\
V3                         & 6.0 cm   & 62.4 ppm          & 62.0 ppm           \\
V3                         & 5.0 cm   & 168.6ppm          & 165.4 ppm           
%\hline
%V1                         & 10 cm    & 28.9 ppm           & 28.8 ppm            \\
%V2                         & 10 cm    & 40.3  ppm         & 39.7 ppm            \\
%V4                         & 10 cm    & 21.3  ppm         & 20.2 ppm            \\
%ADV                      & 10 cm    & 2.8  ppm           & 2.8 ppm            \\
%V3                         & 10 cm    & 19.9  ppm         & 19.4 ppm            \\
%V3                         & 7.5 cm   & 25.2 ppm          & 25.0 ppm            \\
%V3                         & 6.0 cm   & 31.2 ppm          & 31.0 ppm           \\
%V3                         & 5.0 cm   & 84.3 ppm          & 82.7 ppm     
\end{tabular} 
\end{ruledtabular}
\caption{Comparison between losses obtained with the complete simulation and those found with simple round method for different mirrors maps and for different value of the diameter. Results of the quick simulation have been multiplied by two as in the previous case.}
\label{cfr_cq}
\end{center}
\end{table}  

\section{\label{sec:degradation}Squeezing degradation and decoherence estimation}
The quantum noise in an interferometer, normalized with respect to shot noise, can be simply expressed as
\begin{equation}\label{raflos2}
N(\Omega) = 1+ K^{2} (\Omega)
\end{equation} 

where $K $ is the optomechanical coupling defined in Eq.\ref{k}.\

In an ideal system, the injection of a frequency dependent squeezed vacuum from the interferometer dark port will reduce the quantum noise to 
\begin{equation}\label{raflos3}
N (\Omega)= e^{-2\sigma}(1+ K^{2}(\Omega))
\end{equation} 

where $\sigma$ is connected with the squeezing magnitude usually expressed in decibel by $\sigma_{\no{dB}} = -20 \, \sigma\times \log_{10} e$.

In a real setup, two factors prevent this optimal noise reduction: first, optical losses will decrease the squeezing level, introducing a non-squeezed vacuum; second, fluctuations of the squeezing angle will preclude an optimal rotation of the squeezed state.

In \cite{sq2} a detailed analysis of several of these mechanisms has been applied to a 16 m filter cavity, with round trip losses of 1 ppm/m, which is considered a possible short-term solution for Advanced LIGO \cite{sq1}. The same analysis is performed here for the 300 m filter cavity, with round trip losses of 80 ppm, corresponding to a conservative estimate for  Virgo-class quality mirrors. To ease the comparisons between the two cases, the numerical values for other sources of squeezing degradation are exactly the same of \cite{sq2}, as reported in Tab. \ref{val}.

\begin{figure}
	\centering{\includegraphics[width=0.5\textwidth]{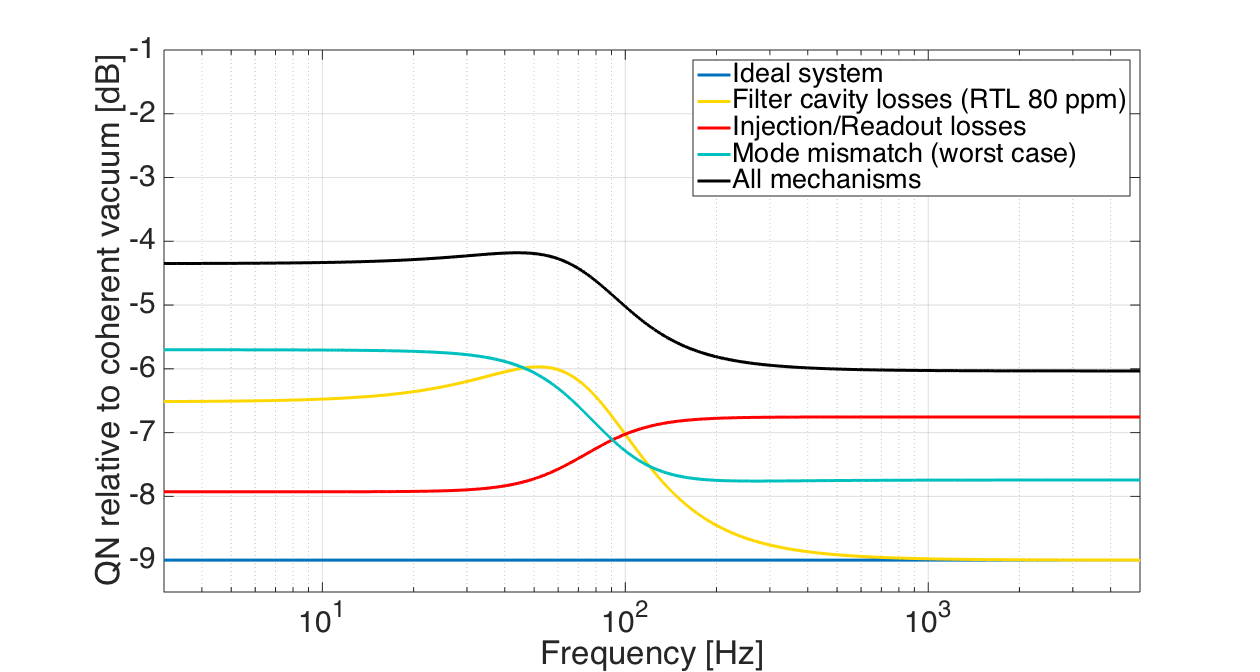}}
	\caption{\textit{Squeezing degradation budget}. Quantum noise relative to coherent vacuum in the signal quadrature for a ideal system (blue curve) is compared with the one obtained taking into account degradation mechanisms (one by one).}
	\label{almec}
\end{figure}

Figure \ref{almec} represents the \textit{squeezing degradation budget} for the 300 m filter cavity. It shows the ratio between the quantum noise of a dual recycled Fabry-P\'erot  Michelson interferometer without squeezing  (for instance, initial KAGRA) and the quantum noise in the presence of frequency dependent-squeezing. An initial realistic squeezing level of 9 dB has been considered and the various degradation mechanisms have been taken into account separately and combined (black curve). In this analysis, as for \cite{sq2}, the contribution of the interferometer losses has been neglected. 
The squeezing degradation mechanisms considered are (for a detailed description see  \cite{sq2}): 

\subsubsection*{ Filter cavity losses}
The optical losses in the filter cavity spoil the squeezing in two ways: firstly, they corrupt squeezing with anti-squeezing, mixing the two squeezing quadratures. Secondly, they mix standard vacuum with squeezed vacuum. For a 16 m cavity, the filter cavity losses represent the main contribution to squeezing degradation up to 100-200 Hz.  Since this effect depends on the round trip losses per meter\cite{Khalili1}, it is considerably reduced in a 300 m cavity with  round trip losses of 80 ppm, and at low frequencies becomes comparable with that of the mode mismatching. For this reason it is not extremely useful to further reduce filter cavity losses by increasing the mirror quality (for example using Advanced Virgo-class mirrors), unless mode matching is substantially improved.

%A comparison of the squeezing deterioration due to losses in the two cavities is shown in Figure \ref{2C}

\subsubsection*{Injection and readout losses}

Injection losses, $\Lambda^{2}_{\mathrm{inj}}$, (caused  by scattering, absorption and imperfections in the optics) and readout losses, $\Lambda^{2}_{\mathrm{ro}}$, (from the interferometer to the readout, including the photodetector quantum efficiency) cause a squeezing degradation by mixing ordinary vacuum with squeezed vacuum. Being independent of the cavity length, their impact does not change with a longer cavity. This mechanism is the dominating source above 100-200 Hz, assuming a losses value of $5\%$ both for injection and readout losses. In this region, quantum noise is by far the limiting noise: a reduction in injection/readout losses will lead to a consistent improvement of the detector sensitivity

\subsubsection*{ Mode mismatching}
Following the analysis shown in \cite{sq2}, a squeezing degradation is also determined  by an imperfect mode matching between the squeezed field and the cavity mode ($\Lambda^{2}_{\mathrm{mmFC}}$), and between the cavity mode and the local oscillator ($\Lambda^{2}_{\mathrm{mmLO}}$). This mismatching allows part of the field to bypass the cavity without experiencing frequency rotation, and are also a source of losses. 

The magnitude of the mode mismatching can be easily measured. Nevertheless, the amplitude of the squeezing degradation depends of an arbitrary phase which is difficult to quantify . For this reason, in Fig.\ref{recapfig} we have shown the worst case scenario. 
 
As in the case of injection and readout losses, this effect does not depend on the filter cavity length and, for a 300 m filter cavity with a RTL of about 80 ppm, it is comparable with the degradation due to filter cavity losses. The estimation is done assuming a mismatch of  $2\%$ between squeezed injected field and filter cavity modes and a mismatch of $5\%$ between injected field and the local oscillator. 

\subsubsection*{ Frequency dependent Phase Noise}

A length noise of the filter cavity (possibly caused by seismic noise or control noise) $\delta L$ results in a shift of the optimal detuning $\delta\Delta\omega_{\no{fc}}$ given by:

	\begin{equation}\label{raflos4}
	\delta\Delta \omega_{\no{fc}} = \frac{\omega_{0}}{L}\delta L
	\end{equation} 
	
This compromises an optimal squeezing quadrature rotation, limiting the achievable noise reduction.
In a 300 m cavity, assuming  a $\delta L$ (RMS) of about 0.3 pm, this effect becomes completely negligible with respect to other mechanisms. 

 \begin{table}[!ht]
 \begin{center}    
 \begin{ruledtabular}
 \begin{tabular}{llr}

Parameter                                                                                 & Symbol                                          & Value \\
\hline

Filter cavity losses                                                                   &$ \Lambda^{2}_{\mathrm{rt}}$                                & $80$ ppm   \\
Injection losses                                                                       & $\Lambda^{2}_{\mathrm{inj}}$                                & $5\% $   \\
Readout losses                                                                       & $\Lambda^{2}_{\mathrm{ro}}$                                & $5\% $            \\
Mode-mismatch squeezer-filter cavity                                    & $\Lambda^{2}_{\mathrm{mmFC}} $                        & $2\% $    \\
Mode-mismatch squeezer-local oscillator                               & $\Lambda^{2}_{\mathrm{mmLO}}$                         & $5\% $   \\
Filter cavity length noise (RMS)                                             & $\delta L_{\mathrm{fc}}$                                                             &$0.3$ pm \\
Injected squeezing                                                                 & $\sigma^{2}_{\mathrm{dB}}$                  & $9\,\mathrm{dB}  $  \\

\end{tabular}
\end{ruledtabular}
\caption{Parameters used in the estimation of squeezing degradation.}
\label{val}
\end{center}
\end{table}

\section{\label{sec: improvement} Improvement in KAGRA sensitivity}

\begin{figure}
	\centering{\includegraphics[width=0.53\textwidth]{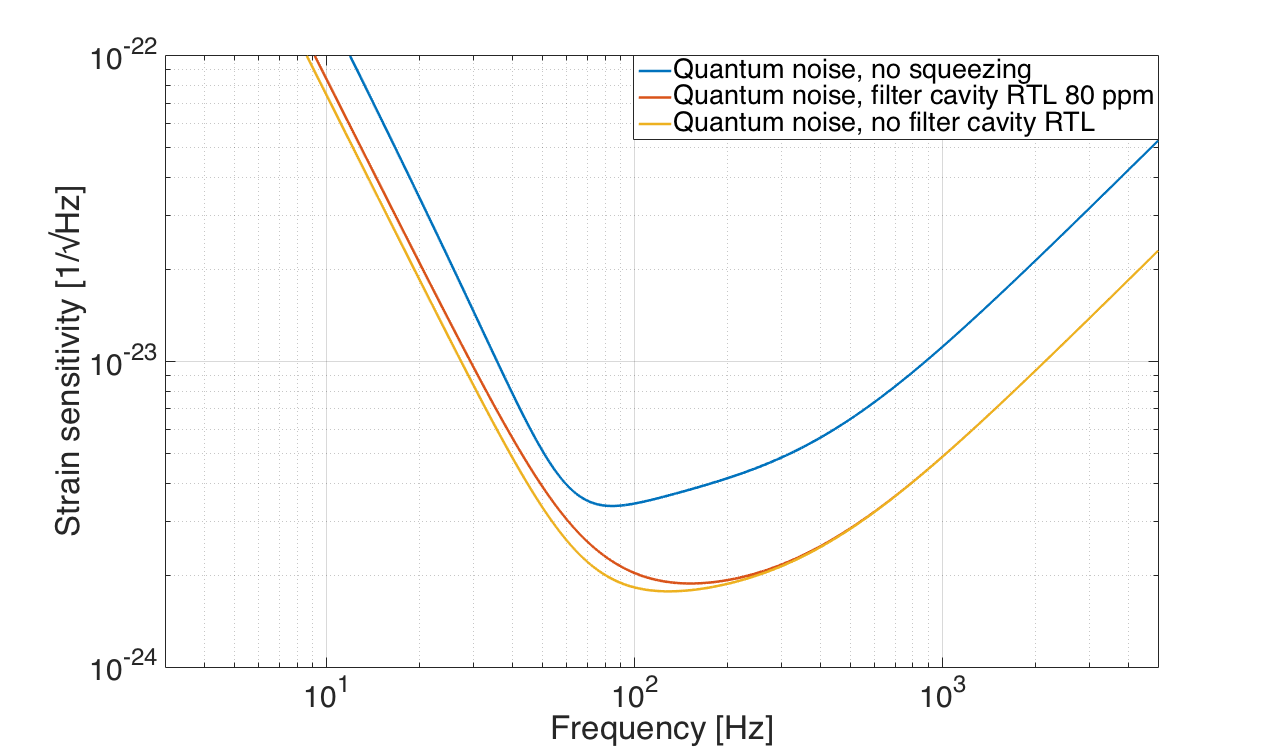}}
	\caption{Quantum noise for KAGRA without squeezing, compared with the quantum noise using 9 dB frequency-dependent squeezing, both in the ideal system and when all the degradation mechanisms are taken into account. We remark that quantum noise without squeezing is relative to official KAGRA sensitivity using a configuration with a homodyne detection angle of 121.8$^{\circ}$, while quantum noise in the presence of squeezing uses a standard homodyne angle of 90$^{\circ}$.}
	\label{rep1}
\end{figure}

This estimation of the achievable level of frequency-dependent squeezing using a 300 m filter cavity can be used to quantify the related improvement in KAGRA sensitivity.  In Fig. \ref{rep1} the quantum noise for KAGRA, without squeezing, is compared with the quantum noise obtained using 9 dB of  frequency-dependent squeezed light, when all  degradation mechanisms previously described are taken into account. We considered both the case of a filter cavity with round trip losses of 80 ppm and that of a perfect filter cavity. The comparison shows that no major improvements can be obtained by reducing round trip losses under the level of $\sim 80$ ppm since their effect becomes comparable with that of other degradation mechanisms.
%in the broadband configuration (tuned signal recycling cavity) and with standard detection angle $\zeta= \pi/2$ 

\begin{figure}
	\centering{\includegraphics[width=0.5\textwidth]{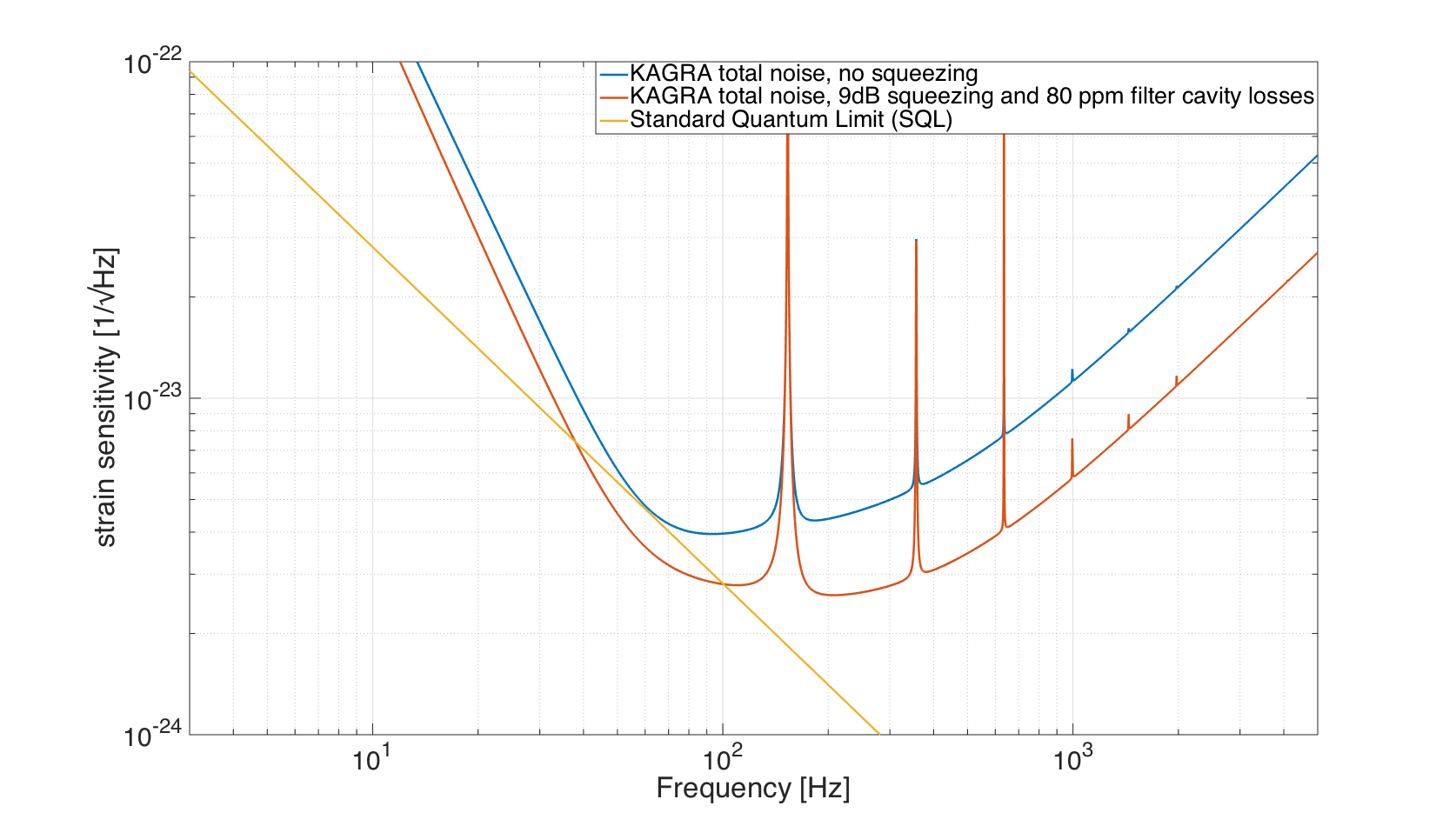}}
	\caption{Improvement in KAGRA sensitivity using 9 dB frequency-dependent squeezing, considering lossy cavity and other degradation mechanisms.}
	\label{rep2}
\end{figure}

\begin{figure}
	\centering{\includegraphics[width=0.5\textwidth]{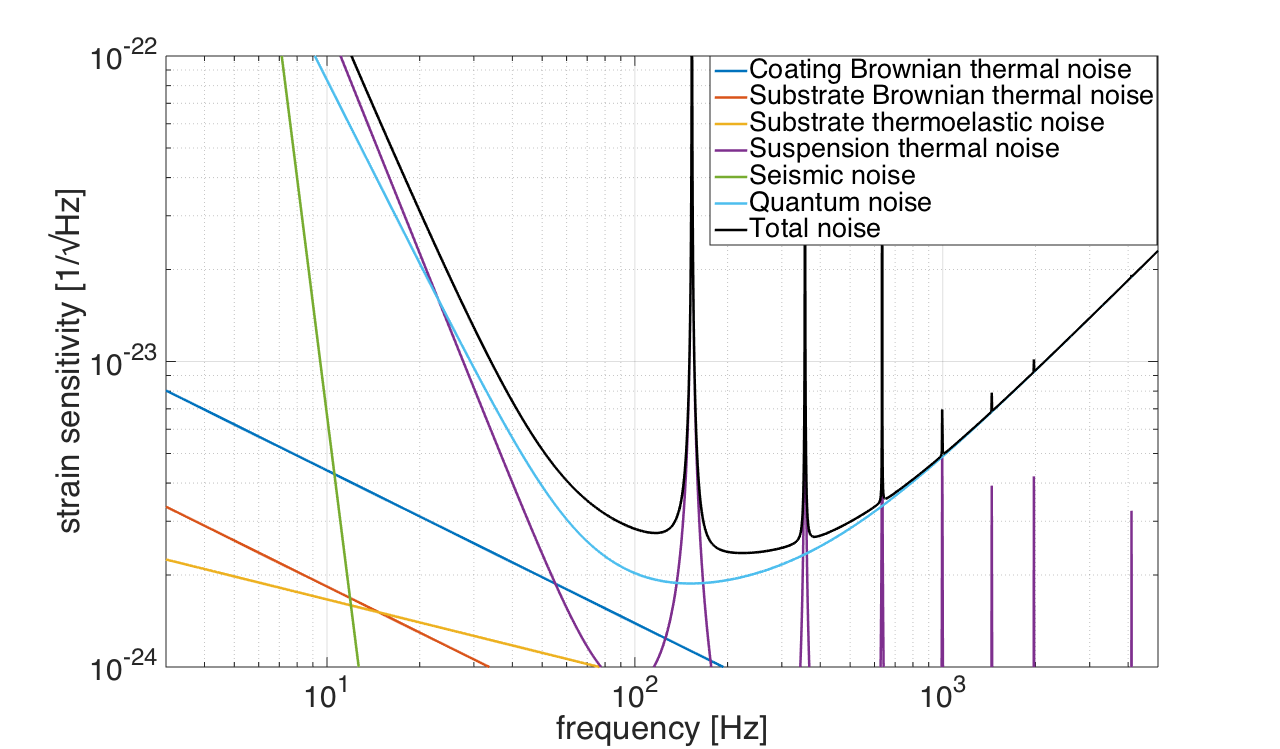}}
	\caption{The quantum noise in the presence of  9 dB squeezing and lossy system (filter cavity RTL 80 ppm) is compared with the other noise sources. Note that below 100 Hz, the contribution from thermal noise would prevent an improvement in the total sensitivity even in case a further reduction of quantum noise.}
	\label{rep3}
\end{figure}

Fig. \ref{rep2} shows the improvement in KAGRA sensitivity using 9 dB frequency-dependent squeezing, considering a filter cavity with round trip losses of 80 ppm and other degradation mechanisms. We remark that the use of squeezing allows us to reach a sensitivity beyond the standard quantum limit around $70$ Hz. In Fig. \ref{rep3} the quantum noise of a realistic lossy system is shown along with other KAGRA noise sources. We observe how a reduction in the only quantum noise will result in an improvement in almost the whole KAGRA observation bandwidth. 

We also remark that:

\begin{itemize}
	\item In the frequency region below 100 Hz, thermal noise is close to quantum noise and there is little to be gained from a further  significant reduction in quantum noise (obtainable by improving mismatching, injection/readout losses and decreasing losses in the filter cavity).
	\item In the frequency region above 100 Hz, which is dominated by quantum noise, there is much more room for improvements. However, a reduction of the filter cavity optical losses would not provide a higher squeezing level, since in this region the squeezing degradation is mainly caused by injection and readout losses. 
\end{itemize}

This suggests that a further reduction of the filter cavity losses would not be necessary, while a major benefit can be obtained by protecting squeezing from losses due to  injection, detection and mismatching.

\section{Summary}

We have presented the optical design of a 300 m filter cavity used for the KAGRA experiment. The cavity will be tested at NAOJ, inside the TAMA infrastructure. 
First, given KAGRA parameters, we have found the detuning and finesse values needed to produce an optimal rotation of the squeezing angle. In order to find the surface quality needed for the filter cavity, we have computed the round trip losses using a numerical simulation with realistic mirror maps. The test bench maps used are  those of Virgo and Advanced Virgo. 

An analysis of the squeezing deterioration mechanism shows that for a round trip loss value of 80 ppm, corresponding to a conservative estimate for Virgo-class mirrors, the squeezing degradation due to the filter cavity are comparable or lower than contributions from the cavity mismatching and injection/detection losses. Moreover  frequency-dependent phase noise becomes completely negligible.

We then computed the improvements in KAGRA sensitivity with a 9 dB frequency dependent squeezing and realistic loss mechanisms. We found a sensitivity improvement in almost the full KAGRA signal bandwidth, with a factor $\sim 2$ at high frequency. As already mentioned, the interferometer internal losses have been neglected in this analysis. In order to have a more precise estimation of the sensitivity improvement achievable using frequency dependent squeezing, the contribution of these losses should be estimated. 

We have also shown that, without an improvement in other noise sources, a further reduction of the filter cavity losses will not lead to a considerable improvement in the detector sensitivity. 

The results of this work can be easily extended to any gravitational-wave interferometric detector planning the use of a 100-m class filter cavity, and in particular for a medium term upgrade of Advanced Virgo and for Einstein Telescope.

\section*{Acknowledgements}

We would like to thank Lisa Barsotti, John Miller and Matteo Tacca for useful discussions. This work was supported by the JSPS Grant-in-Aid for Scientific Research
the JSPS Core-to-Core Program, A. Advanced Research Networks and
the European Commission under the Framework Program 7 (FP7) 'People', project ELiTES (grant agreement 295153). E.C. is supported by the European Gravitational Observatory.

\end{document}